\documentclass[twocolumn,showpacs,aip,apl,preprintnumbers,amsmath,amssymb,superscriptaddress,reprint]{revtex4-1}
\usepackage{graphicx}
\usepackage{enumitem}
\usepackage{booktabs}
\newcommand{\ra}[1]{\renewcommand{\arraystretch}{#1}}
\makeatletter
\def\namedlabel#1#2{\begingroup #2%
	\def\@currentlabel{#2}%
	\phantomsection\label{#1}
\endgroup}
\makeatother
\renewcommand {\t} {\tau}                  
\newcommand {\al} {\alpha}                       
\newcommand{\be}{\begin{equation}}              
\newcommand{\ee}[1]{\label{#1} \end{equation}}  
\newcommand{\bee}{\begin{eqnarray}}             
\newcommand{\eee}{\end{eqnarray}}               

\newcommand{\bracket}[1]{\left\lbrack#1\right\rbrack}

\begin{document}
\title{Highly-complex optical signal generation using electro-optical
systems with non-linear, non-invertible transmission functions}
\author{Jos\'e~J.~Su\'arez-Vargas}
\email[Email:~]{jjsuarez@ivic.gob.ve}
\altaffiliation[Also affiliated at: ]{The ``Abdus Salam" International Centre
for Theoretical Physics, Trieste, Italy}
\affiliation{Centro de F\'{i}sica, Instituto Venezolano de Investigaciones
Cient\'{i}ficas, km. 11 Carretera Panamericana, Caracas 1020-A, Venezuela}
\author{Bicky A. M\'arquez}
\affiliation{Escuela de F\'{i}sica, Facultad de Ciencias, Universidad Central
de Venezuela, Caracas, Venezuela}
\author{Jorge A. Gonz\'alez}
\affiliation{Centro de F\'{i}sica, Instituto Venezolano de Investigaciones
Cient\'{i}ficas, km. 11 Carretera Panamericana, Caracas 1020-A, Venezuela}
\date{\today}
\begin{abstract}
We present a scheme whereby a static non-linear, non-invertible
transmission function performed by the electro-optic Mach-Zehnder
modulator produces highly complex optical chaos. The scheme allows the
deterministic transformation of low-dimensional band-limited chaotic
signals into much higher-dimensional structures with broadband spectra
and without using any delay elements or feedback. Standard benchmark
tests show that all the considered complexity indices are highly
increased due to this transformation in a controlled fashion. This
mechanism allows the design of simple optoelectronic delayed oscillators
with extremely complex chaotic output.
\end{abstract}
\maketitle

The generation of complex signals with broadband spectra is extremely
important in diverse scientific and engineering applications. In
particular chaotic systems have been implemented for this purpose using electrical
and optical circuits. The construction of chaos-based encryption methods and
the versatility of synchronizing chaotic systems have sparked a multitude of
applications for electro-optical communication systems in the last two
decades\cite{Pecora:90}.

The well-known low-dimensional chaotic oscillators (Lorenz, R\"ossler, Chua,
etc.) have been used for studying important dynamic properties like
bifurcation, synchronization and spectral structures, but they lack performance
when used for designing highly secure communication systems. Therefore it has
been necessary to use other schemes for producing more complex chaotic
dynamics. Ikeda\cite{Ikeda:79} introduced a system able to generate
high-dimensional \emph{hyperchaos} (multiple positive Lyapunov exponents)
that has been used as a model of electro-optical systems with delayed
feedback\cite{Neyer:82a}. These systems have shown a wide range of dynamics
from bifurcations, multi-stability and chaotic breathers\cite{Peil:09} in
delayed optoelectronic oscillators (DOO) using the Mach-Zehnder modulator
(MZM) as the source of nonlinearity. With these setups some advanced schemes
of hyperchaos synchronization and optical communication
systems\cite{Goedgebuer:02} were developed.

In this Letter we show a method to generate highly complex chaotic signals based on a robust theory. We
show that using the MZM as a nonlinear static transformation, we can strongly
increase in a controlled way the dimension and other complexity measures of the output signals, including an enhancement of the bandwidth. We differentiate this scheme from others
producing complex signals in that the source of the complexity is not given by
any time delays, but by the topological properties of the transforming function.
As the MZM is already an integral component of advanced chaotic communication
schemes, our mechanism allows to improve the level of security and
information-carrying capacity provided by common hyperchaotic systems.

\begin{figure}
\includegraphics*[scale=0.3,angle=270]{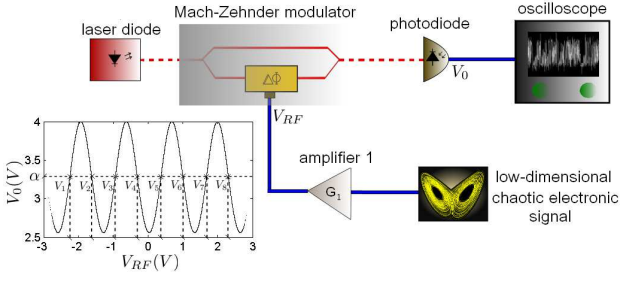}
\caption{\label{schem1_fig1} Schematic diagram of the static non-linear,
non-invertible system based on the MZM transmission function.}
\end{figure}

The theoretical foundation is based on the observation that some
explicit deterministic functions produce random sequences of
values\cite{Gonzalez:97}. These functions result from generalizing the
solutions of standard chaotic mappings. The randomness here is defined in the
sense of the impossibility of writing future values of the sequences as
single-valued relationships with past values (and viceversa). For instance the
function
\begin{equation}
\label{sine1}
\begin{array}{c c l}
X_n&=&\sin^2(\theta\pi z^n),
\end{array}
\end{equation}
where $z$ and $\theta$ are real parameters, is the solution of the
logistic map, $X_{n+1}=4X_n(1-X_n)$, when $z=2$.  However, it was
demonstrated\cite{Gonzalez:01} that the sequences generated by this
function cannot be written as a single-valued map of the form
$X_{n+1} = f (X_n,X_{n-1}, . . . ,X_{n-r+1})$, when $z$ is not integer.
Using this important finding we propose as model of our physical implementation
the system\cite{Gonzalez:02b}:
\begin{equation}
\label{gen_mech}
\begin{array}{c c l}
X(t)&=&h(f(t)),
\end{array}
\end{equation}
in analogy with the structure of Eq. (\ref{sine1}). Deeper analyses have
shown\cite{Gonzalez:01,Gonzalez:02b}
that the necessary conditions to produce complex dynamics in real physical
systems using Eq. (\ref{gen_mech}) are:
\begin{description}[labelwidth=1.5cm]
\item[\namedlabel{itm:rule1}{Condition (i)}]$f(t)$ should be a
non-periodic oscillating function with repeating intervals of finite
growing behavior.
\item[\namedlabel{itm:rule2}{Condition (ii)}]$h(v)$ should be a non-invertible
function with many extrema such that $h(v)=\alpha$ possesses several solutions, ${v_1,v_2,...}$, for $v$.
\end{description}
For the physical implementation of this scheme we chose for $f(t)$ a chaotic
signal generated by a low-dimensional autonomous system. On the other
hand, function $h(v)$ will be implemented by the MZM, which intrinsically
produces the most ideal non-linear non-invertible function, a sinusoidal
transformation. In Fig. \ref{schem1_fig1} the experimental setup shows a laser
diode (1550 nm) coupled through mono-mode optical fiber into the MZM.
The well-known structure of the MZM transmission is a non-linear function
of the applied voltage to the radio-frequency input ($V_{RF}$),
(half-wave voltage $V_{\pi}=5.2$ V), and a DC voltage $V_B$ to set the
operating point on the transmission curve. Then the transmission can be
written as:
\begin{equation}
\label{trans1}
\begin{array}{c c l}
P_o(V_{RF})&=&\eta P_i\cos^2\bracket{\frac{\pi V_{RF}}{2V_\pi}+\phi_o},
\end{array}
\end{equation}
where $P_o$ is the modulated output power, $P_i$ is the power of the
light entering the modulator, $\eta$ is a parameter accounting
for the losses inside the device due to couplings and dissipations and
$\phi_o$ is the bias term depending on $V_B$. The modulated light is
coupled through another small section of mono-mode fiber into the
photodetector (PIN diode) whose output voltage is sent to an oscilloscope for
recording. If in Eq. (\ref{trans1}), we let $P_i$ constant, set $\phi_o=0$ without
losing generality, and let $V_{RF}$ take a chaotic signal we obtain a
topologically equivalent expression to Eq. (\ref{gen_mech}). Now, in order to
ensure that the output will be of higher complexity than the input we have to
fulfill \textbf{\ref{itm:rule2}}. This means that the amplitude of the input
has to span several lobes of the non-invertible transmission of Eq.
(\ref{trans1}). In the setup of Fig. \ref{schem1_fig1} we use an amplifier
of variable gain ($G_1$) that controls the magnitude of the input before
entering into the MZM, and therefore we can control the number of spanned lobes
of the nonlinear transmission. In the insert of Fig. (\ref{schem1_fig1}), we
amplified the input such that four lobes of the transmission were spanned, or
equivalently, a single output value, $\al$, has a possibility of being produced
by eight different input voltages $v_1, v_2, ...,v_8$. This is the source of the
indetermination and statistical independence of the output dynamics. This
parameter $G_1$ will give us another degree of freedom to study the
relationship between the number of lobes of the transformation and the
complexity of the output signal. This observation is key in the production of
highly complex signals using the scheme of this Letter.

As a first application we use some well-known low-dimensional chaotic
oscillators. The Chua, R\"{o}ssler, Lorenz, and Duffing systems were implemented
as electronic circuits\cite{Kennedy:92}.
\begin{figure}
\includegraphics*[scale=0.4]{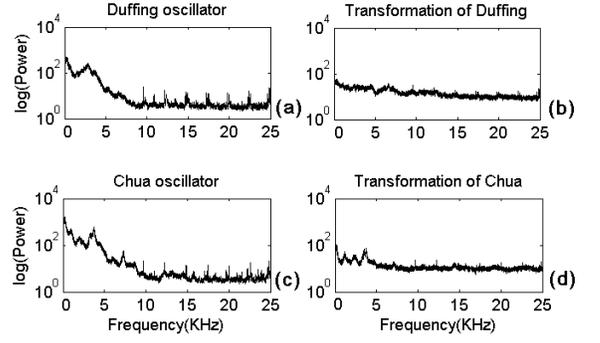}
\caption{\label{FFT_fig2}Spectral characteristics of the chaotic signals,
before \textbf{(a)} and \textbf{(c)}, and after the transformation
\textbf{(b)} and \textbf{(d)}. For Duffing (upper) and Chua's (lower)
oscillators.}
\end{figure}
\begin{figure}
\includegraphics*[scale=0.4]{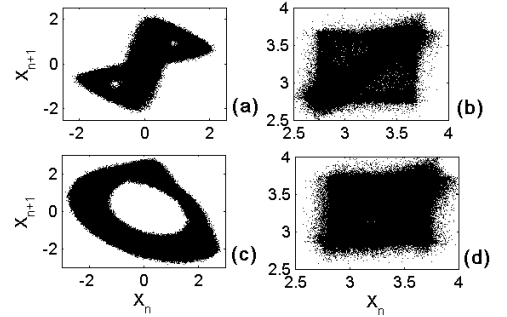}
\caption{\label{FRM_fig3}First return maps of the chaotic signals, before
\textbf{(a)} and \textbf{(c)}, and after the transformation \textbf{(b)} and
\textbf{(d)}. For Lorenz (upper) and R\"ossler (lower) oscillators.}
\end{figure}
One component of each oscillator was amplified with the variable gain $G_1$
(our complexity control parameter) and inserted into the MZM. In Fig.
(\ref{FFT_fig2}), we show the spectral characteristic using the fast fourier
transform (FFT) and first-return maps (FRM), comparing the input signals before
and after the transformation. Interestingly the first observation is that the
FFT of the transformed signals have a flattened noise-like spectra. The
transformation increased their bandwidth, meaning that the original linear
correlations in the chaotic signals are lost due to the transformation. This is
in agreement with the theory from which this procedure is derived. If the
signals lose correlation, the linear dependance between future and past values
is lost, and hence they approach a random process in the sense defined above.
This phenomenon is shown specifically for the chaotic dynamics of Duffing (Fig. (\ref{FFT_fig2}-a,b)), and Chua's (Fig. (\ref{FFT_fig2}-c,d)) circuits. We also
show how the dimension, the ergodicity and uniform distribution of trajectories
of the R\"ossler Fig. (\ref{FRM_fig3}-c) and Lorenz Fig. (\ref{FRM_fig3}-d)
dynamics are increased in the two-dimensional projections given by the FRMs.
The conclusions gathered from these figures can be buttressed by a quantitative
estimation of important parameters that qualify the complexity of chaotic
signals. There are several approaches to classify and study complexity of time
series based on information theory and nonlinear analysis\cite{Kodba:05}. We
use the very robust and professional free software package
TISEAN\cite{Hegger:99} to obtain from the experimental time series the false
nearest neighbor dimension, the maximal Lyapunov exponents and the mutual
information. We chose these particular complexity indices because they are
invariant measures and provide information about the underlaying dynamics
rather than just simple statistics depending on the particular realizations
used to calculate them\cite{Kantz:97}. The results of the calculations are
compiled in Table (\ref{table1}).
\begin{table}
\caption{\label{table1}Complexity indices calculated from experimental
time series before and after the transformation of the system of Fig.
(\ref{schem1_fig1}): False-neighbors dimension (FND), maximal Lyapunov
exponent ($\lambda$) and mutual information (MI).}
\ra{1.0}
\begin{tabular}{@{}ccrrcrrrcrcrr@{}}\toprule
\hline
& \multicolumn{2}{c}{FND} &  & \multicolumn{2}{c}{$\lambda$} &
 & \multicolumn{2}{c}{MI} \\
\cmidrule{2-3} \cmidrule{4-5} \cmidrule{6-8}
\hline
Attractor&before&after&&before&after&&before&after\\\midrule
\hline
R\"ossler &3& 36 && $0.05$ & $2.8$ && 46432.7 & 4.0 \\
Chua      &3& 28 && $0.4$ & $1.4$  && 174.3 & 6.0 \\
Duffing   &3& 31 && $0.6$ & $1.5$ && 146.8 & 4.0 \\
Lorenz    &3& 5  && $0.8$ & $1.7$ && 201.0 & 105.7 \\
\hline
\end{tabular}
\end{table}
We can see that for every attractor transformed the dimension estimated by the
false nearest neighbors (FND) method is strongly increased, obtaining the
highest increment for the R\"ossler attractor. However there was a weak
increment in dimension for the Lorenz signal, which is due to the fact that the
gain $G_1$ used in this case allowed to span only two lobes of the
transformation while for the R\"ossler signal there were six lobes. For Chua
and Duffing the transformation used had four lobes. The maximal Lyapunov
exponent, $\lambda$, that measures the sensibility to initial conditions, or
the \emph{chaoticity} of the attractors are also significantly increased. Specially
notable is the case where the R\"ossler attractor, the most quasi-harmonic
oscillator, increases its maximal Lyapunov exponent 56 times when transformed.
\begin{figure}
\includegraphics*[scale=0.3,angle=270]{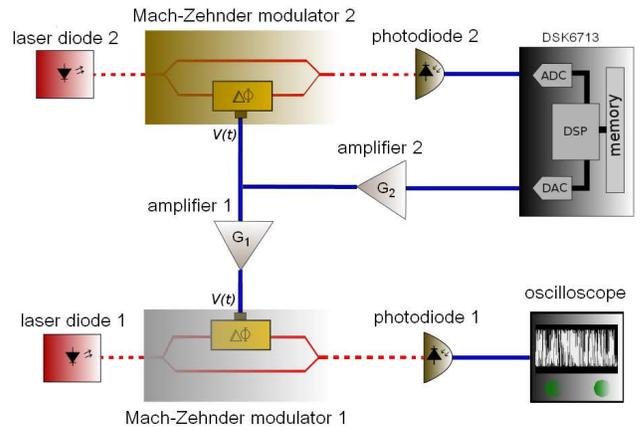}
\caption{\label{schem2_fig4}Schematic diagram of the static non-linear
non-invertible system based on the MZM-1 transmission function, transforming
hyperchaotic signals from the DOO.}
\end{figure}
If the FFT figures above gave us an estimate of the elimination of linear
correlations produced by our transformation, the mutual information (MI),
accounts for non-linear correlations present in the time series\cite{Fraser:86}.
When we calculate the MI on the transformed signals, the value $\t$ for the first zero in the MI is very close to zero, meaning this transformation eliminates almost absolutely
any linear or non-linear correlations present in the original dynamics.

We now turn to the question of the possible limitations of this method. That is,
what happens when the original signal is already very complex and
high-dimensional? Will the transformation produce the above results? To answer
this we focus on delayed-feedback optoelectronic oscillators (DOO) which are
known to generate some of the most complex chaotic behavior\cite{Larger:2005}.
We implement as the source of the chaotic signal to be transformed a
low-frequency DOO as described by Murphy et al.\cite{Murphy:10}. Although this
oscillator is very well described by many of the references of this Letter, we
briefly mention our setup (see Fig. (\ref{schem2_fig4})): A 1550 nm laser diode
(LD-2) with constant power (500 mW) lights through an optical fiber the MZM-2,
the optical output of this device is converted to electrical voltage by a PIN
photodetector (PD-2). Then this voltage is inserted into a digital signal
processor board (DSP) TMDSDSK6713 of Texas Instruments. Inside the board we
implement digital filtering to adjust the bandwidth of the chaotic signal in
the range 100 Hz--10 KHz as well as digital delay that we set in 2.6 ms. This
board is designed for audio frequencies with a maximum sampling rate of 96 KHz.
After the signal is delayed and filtered by the DSP, it is converted back to
analogue voltage and amplified by the constant gain $G_2=19.6$. Finally this
signal is inserted into $V_{RF}$ of the MZM-2 ($V_{\pi}=4.5 V$) to close the
loop.
\begin{figure}
\includegraphics*[scale=0.4]{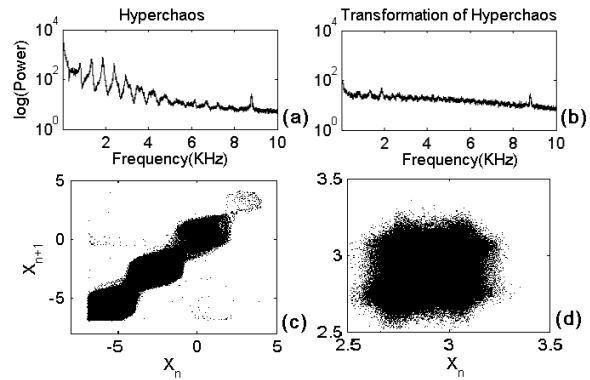}
\caption{\label{FFT2-FRM_fig5}First row: FFT of the hyperchaos generated by the
DOO before (a) and after (b) the transformation. Second row: FRM of the
hyperchaos before (c) and after (d) the transformation.}
\end{figure}
With the parameters set as above, the DOO performs hyperchaos with an FFT
shown in Fig. (\ref{FFT2-FRM_fig5}-a). This signal is fed through the amplifier
with gain $G_1=2$ into the transforming system. We would like to remark that the
subsystems of Fig. (\ref{schem2_fig4}) are two separate and independent entities
with the upper system providing the hyerchaotic signal driving the lower system
that performs the transformation. In the results shown in the first row of Fig. (\ref{FFT2-FRM_fig5}) we plot the FFT of the hyperchaos before (a)
and after (b) the transformation. As in the low-dimensional case, we see that
the correlations present in the original hyperchaos are lost in the
transformation. In the second row we observe FRMs for the hyperchaos before
(c) and after (d) the transformation, the increase of dimension and ergodicity
of orbits is again evident. Additionally and to compare with the
low-dimensional study we obtain the same complexity indices shown in Table
(\ref{table2}).
\begin{table}
\caption{\label{table2}Complexity indices calculated from experimental time
series before and after the transformation of the system of Fig.
(\ref{schem2_fig4}): False neighbors dimension (FND), maximal Lyapunov exponent
($\lambda$) and mutual information (MI).}
\ra{1.3}
\begin{tabular}{@{}rrrrcrrrcrrr@{}}\toprule
\hline
& \multicolumn{2}{c}{FND} &  & \multicolumn{2}{c}{$\lambda$} &
 & \multicolumn{2}{c}{MI}\\
\cmidrule{2-3} \cmidrule{4-6} \cmidrule{7-9}
\hline
Attractor & before & after && before & after && before & after\\\midrule
\hline
Hyperchaos & 35~~~ & $>80$ && $1.5$ & $3.0$ && 66.3 & 6.0\\
\bottomrule
\hline
\end{tabular}
\end{table}
\begin{table}
\caption{\label{table3}KL dimension and entropy for different transformations
of the system of Fig. (\ref{schem2_fig4}): KL dimension (KLD) and Shannon
entropy (H).}
\begin{tabular}{ccc}
\hline
Transformation ($G_1$) & KLD & $H$ \\
\hline
0&9&2.43\\
2&26&4.48\\
3.75&43&4.64\\
5&51&5.45\\
\hline
\end{tabular}
\end{table}
We see that the dimension, although originally high, is strongly increased
up to the limits of calculations of the algorithms and software implementation
given by the TISEAN. Also both the Lyapunov exponent and the mutual information
are again significantly increased, meaning that transforming the hyperchaos
produces a signal \emph{more hyperchaotic} and totally uncorrelated.

As a final test on the functionality of the method we added yet another
rigorous analysis based on the Karhunen-Lo\'eve (KL)
decomposition\cite{Kirby:01} that has been implemented\cite{Franz07} on delayed
electro-optical systems to assess how the time delays affect the complexity of
the signals. This decomposition evaluates the dimension in terms of the number
of principal orthogonal modes needed to reconstruct the signal and computes the
Shannon entropy from the eigenvalue spectra to quantify the dynamic complexity.
In Table \ref{table3} we show the calculations of the KL dimension and the
Shannon entropy for different values of the gain ($G_1$) of the dynamics
generated by system of Fig. (\ref{schem2_fig4}). The first instance corresponds
to the hyperchaos without transformation ($G_1=0$). Then we increase by steps
$G_1$ to apply the transformation with increasing number of lobes from four to
eight. The results show a monotone increase of the KL dimension and entropy,
meaning that the complexity of the signal can indeed be controlled in this way.

In summary we have shown a method to construct deterministically very complex
signals out of simpler inputs, which can be readily implemented in the design
of chaos-based optical communication systems. The signals produced have a wider
bandwidth than those used to generate them and have statistical properties of
random processes with very high values of complexity quantifiers. In
optoelectronic delayed oscillators, the chaotic complexity arises by virtue of
delays in the system, in our scheme the complexity comes determined by the
topological properties of the transformation. Therefore an interesting question
would be the relation arising between time delay and non-invertibility in the
generation of very complex deterministic signals. We hope this work sheds some
light and motivation in this direction.

We thank R. Roy, B. Ravoori and A. Cohen for help in implementing
the delayed optoelectronic oscillator through the ``Hands-on Research on
Complex Systems School" (2009) supported by the ICTP. We also thank W.
Br\"amer of IVIC for helpful discussions and providing equipments.
%
\end{document}